\documentclass[aps, prb, twocolumn, superscriptaddress, amsmath,  tightenlines, longbibliography]{revtex4-1}

\usepackage{dcolumn}
\usepackage{graphicx}
\usepackage{mathrsfs}
\usepackage{subfigure}
\usepackage{booktabs}
\usepackage{amsmath}
\usepackage{physics}
\usepackage{dsfont}
\usepackage{amstext}
\usepackage{amssymb}
\usepackage{amsbsy}
\usepackage{bbm}
\usepackage{amsthm}
\usepackage{graphicx}
\usepackage{color}
\usepackage[colorlinks,citecolor=blue]{hyperref}

\usepackage{url}
\usepackage[colorlinks]{hyperref}
\hypersetup{%
	plainpages=true,
	breaklinks=true,       
	hypertexnames=false,  
	pageanchor=true,
	colorlinks=true,
	linkcolor={blue},
	citecolor={red},
	urlcolor={blue},
	anchorcolor={black}
}

 \makeatletter

\newcommand{\Rmnum}[1]{\expandafter\@slowromancap\romannumeral #1@}
\makeatother

\begin{document}

\title{Non-Hermitian Skin Effect in Periodically-Driven Dissipative Ultracold Atoms}
\author{Zhao-Fan Cai}
\affiliation{School of Physics and Optoelectronics, South China University of Technology,  Guangzhou 510640, China}
\author{Tao Liu}
\email[E-mail: ]{liutao0716@scut.edu.cn}
\affiliation{School of Physics and Optoelectronics, South China University of Technology,  Guangzhou 510640, China}
\author{Zhongmin Yang}
\email[E-mail: ]{yangzm@scut.edu.cn}
\affiliation{School of Physics and Optoelectronics, South China University of Technology, Guangzhou 510640, China}
\affiliation{Research Institute of Future Technology, South China Normal University, Guangzhou 510006, China}
\affiliation{State Key Laboratory of Luminescent Materials and Devices and Institute of Optical Communication Materials, South China University of Technology, Guangzhou 510640, China}

\date{{\small \today}}


\begin{abstract}
	The non-Hermitian skin effect (NHSE), featured by the collapse of bulk-band eigenstates into the localized boundary modes of the systems, is one of most striking properties in the fields of non-Hermitian physics. Unique physical phenomena related to the NHSE have attracted a lot of interest, however, their experimental realizations usually require nonreciprocal hopping, which faces a great challenge in ultracold-atom systems. In this work, we propose to realize the NHSE in  a 1D optical lattice  by periodically-driven ultracold atoms in the presence of staggered atomic loss.  By studying the effective Floquet Hamiltonian in the high-frequency approximation, we reveal the underlying mechanism for the periodic-driving-induced the NHSE. We found that the robust NHSE can be tuned by driving phase, which is manifested by the dynamical localization. Most remarkably, we uncover the periodic-driving-induced critical skin effect for two coupled chains with different driving phases, accompanied by the appearance of size-dependent topological in-gap modes. Our studies provide a feasible way for observing the NHSE and exploring corresponding  unique physical phenomena due to the interplay of non-Hermiticity and many-body statistics in ultracold-atom systems. 
\end{abstract}

\maketitle

\section{Introduction} 

Non-Hermitian systems exhibit many unique physical phenomenon without Hermitian counterparts, which have drawn extensive research interest in recent years \cite{RevModPhys.88.035002,PhysRevLett.116.133903, PhysRevLett.118.040401, PhysRevLett.118.045701,arXiv:1802.07964,El-Ganainy2018,ShunyuYao2018,PhysRevLett.125.126402, PhysRevLett.123.066404, YaoarXiv:1804.04672,PhysRevLett.121.026808,PhysRevLett.122.076801,PhysRevB.98.205417,PhysRevB.100.184314,PhysRevB.102.205423,PhysRevLett.123.170401, PhysRevLett.123.206404,PhysRevLett.123.066405,PhysRevLett.123.206404,PhysRevB.99.201103, PhysRevB.100.054105,PhysRevB.99.235112,Zhao2019,PhysRevX.9.041015,PhysRevLett.124.056802,PhysRevB.102.235151, Li2020,PhysRevB.104.165117, Ashida2020, PhysRevB.102.205118,   PhysRevLett.124.086801, PhysRevLett.125.186802,PhysRevLett.127.196801,Li2021, RevModPhys.93.015005,PhysRevLett.128.223903,  Zhang2022,PhysRevLett.129.093001, Ren2022,PhysRevX.13.021007,PhysRevLett.131.036402,PhysRevLett.131.116601,arXiv:2403.07459, Okuma2023}. In particular, one of their striking physical properties is the emergence of the non-Hermitian skin effect (NHSE) \cite{ ShunyuYao2018, YaoarXiv:1804.04672,PhysRevLett.122.076801,PhysRevLett.123.066404,PhysRevLett.125.126402,  PhysRevLett.121.026808,Li2020,PhysRevB.104.165117, Ashida2020, PhysRevB.102.205118,   PhysRevLett.124.086801, PhysRevLett.125.186802}. By the NHSE,  an extensive number of bulk modes collapse into localized boundary modes in the open boundaries. The NHSE has its intrinsic topological origin associated to the point gap \cite{PhysRevX.9.041015,PhysRevLett.124.086801}, and it can lead to many exotic physical phenomena, e.g., breakdown of conventional Bloch band theory \cite{ShunyuYao2018,PhysRevLett.123.066404,PhysRevLett.125.126402}, nonunitary scaling  of non-Hermitian localization \cite{PhysRevLett.126.166801}, and entanglement phase transitions\cite{PhysRevX.13.021007}.

Although several theoretical proposals have devoted to exploring the NHSEs and several related novel physical phenomena \cite{RevModPhys.93.015005,Lin2023,Okuma2023},   experimental studies remained largely fall behind. The main obstacle, in most cases,  is the challenging requirement of the nonreciprocal hopping in achieving the NHSE for many experimental platforms. At present, most experiments are limited to classical  phononic and optical structures \cite{Weidemann2020,Wang2021,Zhou2023}, and electrical circuits \cite{Helbig2020,Zou2021}.  A more operational approach is to utilize onsite gain and loss or equivalently imbalanced onsite dissipation \cite{PhysRevLett.124.250402, PhysRevLett.125.186802, PhysRevLett.128.223903, PhysRevLett.129.070401,Ren2022}.

Ultracold quantum gases in optical lattices provide a promising platform for studying intriguing quantum physics  due to their high controllability, rich lattice structures and many-body nature \cite{RevModPhys.80.1215, Zhang2018,RevModPhys.91.015005,Schfer2020}, which have been widely utilized to explore topological phases \cite{ Zhang2018,RevModPhys.91.015005} and many-body physics \cite{Schfer2020}. Furthermore, NHSEs have been theoretically proposed  \cite{PhysRevLett.124.250402,  Zhou2022,PhysRevA.106.L061302} and also experimentally observed \cite{PhysRevLett.129.070401, Ren2022} in optical lattices of ultracold atoms. These studies usually rely on complicated lattice structures to achieve the NHSE  assisted by atomic loss. 

Alternatively, we may realize the NHSE  and study the associated physical phenomena in an  experimentally accessible optical lattice of ultracold atoms based on  Floquet engineering. Floquet engineering is a versatile tool by tailoring a system using the  periodic driving, and has produced a wide variety of fascinating  physics in the field of ultracold atoms due to their excellent dynamic control \cite{PhysRevX.4.031027, Goldman2016,RevModPhys.89.011004, Grg2019, Weitenberg2021}.  Most  recently,  loss induced NHSEs have been reported in periodically-driven photonic structure \cite{arxiv.2306.04460} and  acoustic
metamaterial  \cite{arxiv.2306.10000}.  Both proposals are classical settings, and it is natural to achieve the NHSE  in periodically driven ultracold-atom systems, which would offer an exciting opportunity for studying the NHSE  and its interplay with many-body interaction. 
 
In this paper, we theoretically propose to realize the NHSE in periodically-driven dissipative ultracold atoms. By periodically driving the 1D optical lattice of ultracold atoms in the presence of the staggered atomic loss, we observe the NHSE with all the modes localized at boundaries, which can be manifested by dynamical localization.  The direction of the skin-mode localization can be controlled by the driving phase, and characterized by the winding number.  Moreover, by coupling two periodically-driven chains with different driving phases, we study the critical skin effect, with the appearance of size-dependent topological in-gap modes.  

The rest of this paper is organized as follows. In Sec.~\Rmnum{2}(A,B), we present  a 1D optical lattice constructed by dissipative ultracold atoms under the  periodic modulation of onsite potential. When the staggered atomic loss is introduced along the chain, we observe the NHSE. We discuss the effect of disorder and dynamical localization in Sec.~\Rmnum{2}(C) and Sec.~\Rmnum{2}(D). To go further for demonstrating the novel non-Hermitian physics in periodically driven system, we study the critical skin effect in Sec.~\Rmnum{3}.  Finally, we summarize the key findings  in Sec.~\Rmnum{4}.

\section{Periodically-Driven Dissipative Ultracold Atoms} 

\subsection{Model and Floquet Hamiltonian}

\begin{figure}[!tb]
	\centering
	\includegraphics[width=8cm]{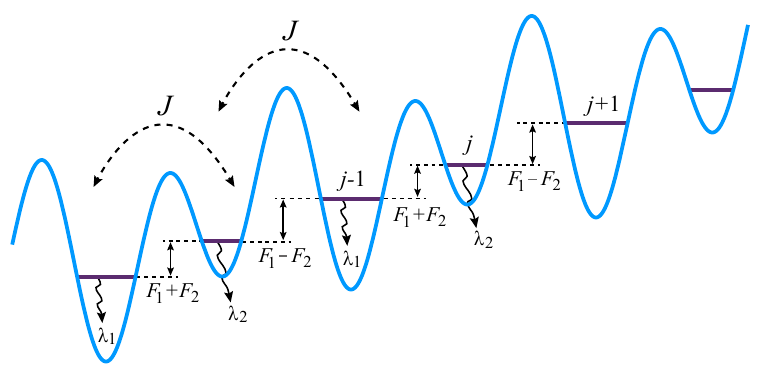}
	\caption{Schematic showing the tilted optical lattice of ultracold atoms with periodic modulation of onsite potential. The bare atomic tunnel coupling is represented by $J$,   $\lambda_1$ and $\lambda_2$ with $\lambda_2  = \lambda_1 + \lambda $ denote the staggered atomic loss at odd and even sites, and onsite potential modulations are expressed as $F_1 = F \cos(\omega t+\phi)$, and $F_2 = F \sin(\omega t)$. }\label{FigScheme}
\end{figure}

To induce NHSE, we consider  periodically-driven dissipative  quantum systems, which can be experimentally realized by using well-developed techniques of Floquet engineering  in ultracold atoms, including the approaches of lattice shaking, laser-assisted tunneling, modulation of external field gradients and a combination of these methods \cite{PhysRevX.4.031027, Goldman2016,RevModPhys.89.011004, Grg2019, Weitenberg2021}.  In 1D dissipative ultracold atoms, by  periodic modulation of onsite potential in 1D optical lattices in a tilted optical lattice (see Fig.~\ref{FigScheme}), we construct the periodically-driven Hamiltonian $\mathcal{H}_\textrm{p}(t) = \mathcal{H}_0 + \mathcal{H}_\textrm{d}(t)$, with
\begin{align}\label{single1}
	\mathcal{H}_0 = -J \sum_{j=1}^{L-1}   \left(c^\dagger_{j+1} c_{j} + \text{H.c.}\right)   - i \lambda \sum_{j=1}^{L/2} n_{2j} - i \lambda_1 \sum_{j=1}^{L} n_j ,
\end{align}
\begin{align}\label{periodically}
	\mathcal{H}_\textrm{d}(t) = \sum_{j=1}^{L} F\left[ \cos(\omega t+\phi) j   +  \sin(\omega t)   \frac{3+(-1)^{j}}{2} \right]  n_{j}.
\end{align}
Here, $\mathcal{H}_0$ is the undriven Hamiltonian in presence of staggered dissipation with    $\lambda_1$ and $\lambda_2=\lambda_1+\lambda$ being atomic loss rate at odd  and  even sites   (see Fig.~\ref{FigScheme}), and $\mathcal{H}_\textrm{d}$ represents periodically-modulated onsite potential in a tilted optical lattice, with its first term being onsite energy offset along the lattice, and its second term being staggered onsite potential. $c^\dagger_{j} $ is the creation operator at the $j$th lattice site,  $n_j = c^\dagger_{j} c_{j}$ is corresponding density operator,   $F$,  $\omega$, and $\phi$ are driving strength, frequency and phase, and $J$ is atomic hopping rate.  The details of experimental proposals are shown   in Appendix \ref{Appendix_AA}.  Without loss of generality, we set $\lambda_1 = 0$ below. 

By performing an unitary transformation $\mathcal{U}(t) = \exp (i\int^{t} \mathcal{H}_\textrm{d}(t') dt' )$ to a rotating
frame of reference  (see Appendix \ref{Appendix_A}),   $\mathcal{H}_\textrm{p}$ is  rewritten as
\begin{align}\label{1Dmodel}
	\mathcal{H}(t) =  &  ~~ \sum_{j}  \left[\left(u c^\dagger_{2j} c_{2j-1} + v c^\dagger_{2j+1} c_{2j} + \text{H.c.}\right) -i \lambda  n_{2j} \right],
\end{align}
where 
\begin{align}\label{1Dmode21}
u = -J e^{i \sqrt{2}\alpha \cos(\frac{\phi}{2} + \frac{\pi}{4}) \sin(\omega t + \frac{\phi}{2} -\frac{\pi}{4})},
\end{align}
\begin{align}\label{1Dmode22}
v =- J e^{i \sqrt{2}\alpha \sin(\frac{\phi}{2} + \frac{\pi}{4}) \cos(\omega t +\frac{\phi}{2}-\frac{\pi}{4})},
\end{align}
with $\alpha = \sqrt{2} F /\omega$.

\begin{figure*}[!tb]
	\centering
	\includegraphics[width=18cm]{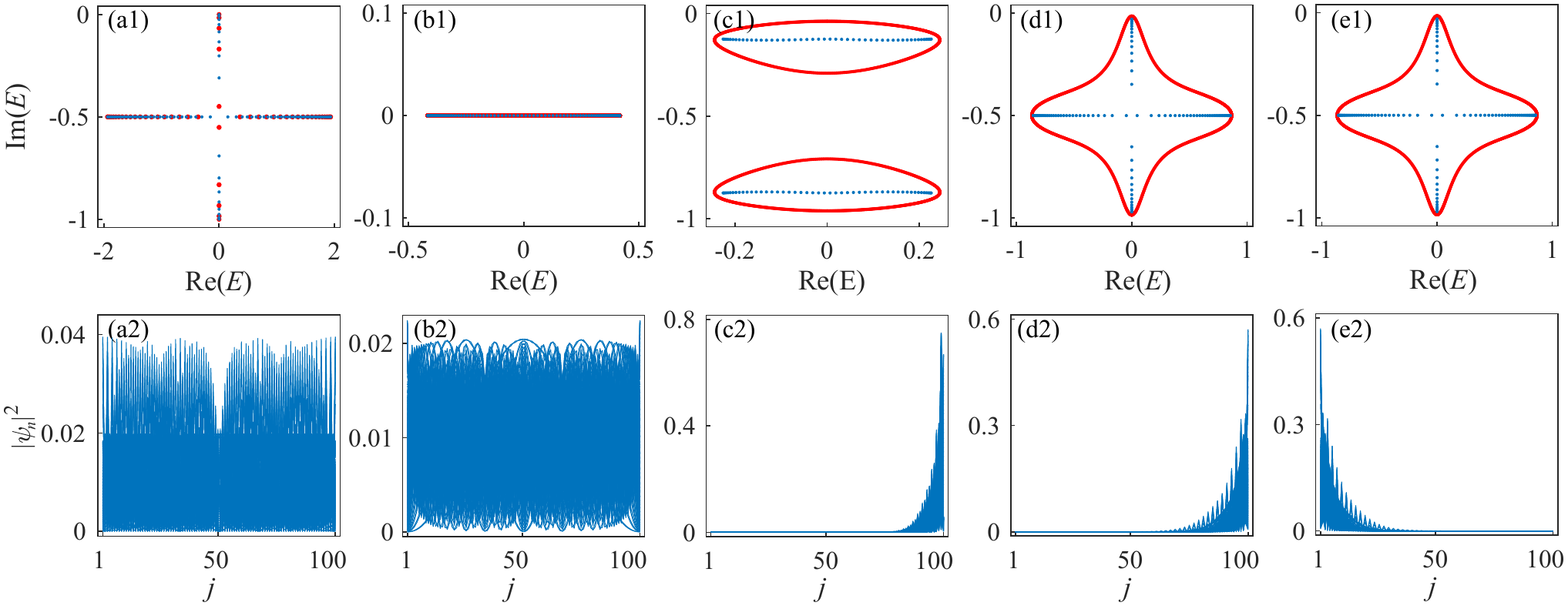}
	\caption{Complex quasienergy spectrum $E_n$ and corresponding density distributions $\abs{\psi_n}^2$ of the  Floquet Hamiltonian $\mathcal{H}_\textrm{F}$. The red and blue dots indicate the quasienergies under PBC and OBC, respectively.
	(a1, a2) Pure dissipative ultracold atoms with  $\alpha=0$, and (b1, b2) the only periodic drive with  $\alpha=2$, $\omega = 2\pi$ and $\lambda =0$.  Periodically-driven dissipative ultracold atoms with (c1, c2)  $\alpha=2$, $\omega = 2\pi$ and $\phi =0$, (d1,d2) $\alpha=1.5$, $\omega=3\pi$ and $\phi =0$, and (e1,e2) $\alpha=1.5$, $\omega = 3\pi$ and $\phi = \pi$. The other parameters used are $\lambda/J = 1.0$, and $L=100$. }\label{energydensity}
\end{figure*}

According to Floquet theorem \cite{Rudner2020}, a time-periodic Hamiltonian $\mathcal{H}(t) = \mathcal{H}(t+T)$, with the driving period $T = 2 \pi /\omega$, is governed  by  Schr$\ddot{\text{o}}$dinger equation $i\partial_t |\psi_n(t)\rangle  =  \mathcal{H}(t) |\psi_n(t)\rangle$. There exists a complete set of orthogonal solutions $|\psi_n(t)\rangle = e^{-i E_n t} |u_n(t)\rangle$, with $|u_n(t)\rangle = |u_n(t+T)\rangle$ and quasienergy $E_n$. In this work, we  are  interested in the stroboscopic dynamics governed by the time-independent effective Floquet Hamiltonian $\mathcal{H}_\textrm{F}$, defined as
\begin{align}\label{UnitaryEvolution}
	U(T)   =     \mathcal{T} e^{-i\int_{0}^{T} \mathcal{H}(t') dt'} = e^{-i\mathcal{H}_\textrm{F} T } ,
\end{align}
where $\mathcal{T}$ is the time-ordering operator, and $U(T) |\psi_n(0)\rangle = e^{-i E_n T} |\psi_n(0)\rangle $.  Due to the non-Hermiticity of $\mathcal{H}(t)$, the Floquet operator $U(T)$ is not unitary, and the quasienergy $E_n$ can be complex. 

\subsection{Non-Hermitian Skin Effect}

To reveal the periodically-driven skin effect for the dissipative ultracold-atom system proposed above, we plot the complex quasienergy spectrum $E_n$ and corresponding density distributions $\abs{\psi_n}^2$ of the  Floquet Hamiltonian $\mathcal{H}_\textrm{F}$, as shown in Fig.~\ref{energydensity}. The  quasienergies  are calculated under both periodic (PBC) and open (OBC) boundary conditions, while density distributions are under the OBC. The complex quasienergy spectrum of the pure dissipative system with $\alpha = 0$ is insensitive to the boundary conditions, and there exists no NHSE [see Figs.~\ref{energydensity}(a1,a2)]. In addition, the only periodically-driven ultracold atoms with $\lambda = 0$ cannot induce the NHSE [see Figs.~\ref{energydensity}(b1,b2)].
 
For the periodically-driven and dissipative ultracold-atom chain in the presence of the staggered loss (i.e., $\alpha \neq 0$ and $\lambda \neq 0$), the complex quasienergy spectrum  is sensitive to boundary conditions  [see Figs.~\ref{energydensity}(c1,d1,e1)]. The periodic drive induces the formation of point gaps, a loop in the complex plane [see red dots in Figs.~\ref{energydensity}(c1,d1,e1)], encircling the open-arc spectrum of OBC (blue dots). Figure  \ref{energydensity}(c2,d2,e2) plots the density distributions $\abs{\psi_n}^2$ of all the Floquet eigenmodes, where all the modes are localized at the boundaries. Moreover, these non-Hermitian skin modes can be localized at either right [see Figs.~\ref{energydensity}(c2,d2)] or left [see Figs.~\ref{energydensity}(e2)] boundaries, determined by the driving phase $\phi$. Therefore, our periodic-modulation approach realizes a tunable 1D NHSE in the dissipative ultracold-atom system.

The NHSE  has its topological origin, which is characterized by the point-gap topology with the topological invariant of winding number \cite{PhysRevLett.124.086801}
\begin{align}\label{windingNumber}
	\mathcal{W}(E_r) = \int_{-\pi}^{\pi}\frac{dk}{2\pi i} \partial_k \ln \det \left[\mathcal{H}_\textrm{F}(k) - E_r\right],
\end{align}
where $\mathcal{H}_\textrm{F}(k)$ is the Floquet Hamiltonian  in momentum space, and $E_r$ is a reference energy point inside the loop of the point gap. $\mathcal{W}(E_r)$ counts the number of times that the complex spectrum of $\mathcal{H}_\textrm{F}(k)$ encircles $E_r$. 

We  analytically calculate $\mathcal{W}(E_r)$ by approximating $\mathcal{H}_\textrm{F}(k)$ using the time-independent effective Hamiltonian $\mathcal{H}_\textrm{eff}(k)$, obtained by the Floquet-Magnus expansion in the high-frequency approximation \cite{Bukov2015, Eckardt2015, PhysRevB.93.144307}.  For $\omega \gg J, F$, the effective Hamiltonian $\mathcal{H}_\textrm{eff}(k)$ (see details in Appendix \ref{Appendix_B}) is derived as 
\begin{align}\label{Hamileffkfinal1}
	\mathcal{H}_\textrm{eff}(k) = & - \sum_{k} \left[JJ_{0}(\alpha) \left(1+e^{-i k}\right) c^\dagger_{k,A} c_{k,B} + \textrm{H.c.} \right] \notag \\
	& - s\frac{2J^2 J_{-1}^2(\alpha)}{\omega} \sum_{k} \sin(k) \left( c^\dagger_{k,A} c_{k,A} - c^\dagger_{k,B} c_{k,B} \right) \notag \\
	&   -i\lambda \sum_{k} c^\dagger_{k,B} c_{k,B},
\end{align}
where $A$ and $B$ denote two sublattice sites in the unit cell due to the staggered loss, $J_m(\alpha)$ is the Bessel function of the first kind, and $s = 1 ~(-1)$ corresponding to $\phi = 0 ~(\pi)$, corresponding to the rightward (leftward) localization. The winding number is calculated as $\mathcal{W} = -1$ ($\mathcal{W} = 1$) for $\phi = 0 ~(\pi)$.

\begin{figure}[! b]
	\centering
	\includegraphics[width=8.4cm]{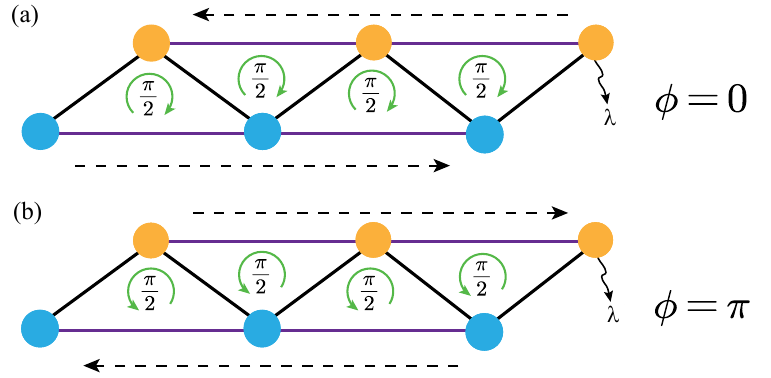}
	\caption{Schematic showing the 1D  zigzag lattice described by Eq.~(\ref{HamilefffinalSM21}). The even sites (yellow filled circles) are dissipated by $\lambda=\lambda_2-\lambda_1$.  A $\pi/2$ magnetic flux threads through each triangular plaquette in  clockwise and counter-clockwise directions for (a) $\phi=0$  and (b) $\phi=\pi$  due to the complex next-nearest-neighbor hopping  in Eq.~(\ref{HamilefffinalSM21}). Such a nonzero flux leads to opposite particle transports along the odd and even lattice sites, indicated by the dashed arrows.  }\label{magneticflux1D} 
\end{figure}

To further elucidate the mechanism of Floquet induced NHSE considered here, we write down the real-space effective   Hamiltonian in the high-frequency expansion (see details in Appendix \ref{Appendix_B}) as
\begin{align}\label{HamilefffinalSM21}
	\mathcal{H}_\textrm{eff} = &- \sum_{j} \left[JJ_{0}(\alpha)c^\dagger_{2j} c_{2j-1} + JJ_{0}(\alpha) c^\dagger_{2j+1} c_{2j} + \textrm{H.c.} \right]   \notag \\
	& - \sum_{j} \left[i s\frac{2J^2 J_{-1}^2(\alpha)}{\omega} c^\dagger_{2j+1} c_{2j-1} + \textrm{H.c.}\right]    \notag \\
	& + \sum_{j} \left[is\frac{2J^2 J_{-1}^2(\alpha)}{\omega} c^\dagger_{2j+2} c_{2j} + \textrm{H.c.}\right]  \notag \\
	& -i\lambda \sum_{j} c^\dagger_{2j} c_{2j}.
\end{align}

The Hamiltonian $\mathcal{H}_\textrm{eff}$ in Eq.~(\ref{HamilefffinalSM21}) describes an effective 1D lattice in the presence of both nearest-neighbor and next-nearest-neighbor hoppings, forming a zigzag lattice [see Fig.~\ref{magneticflux1D}]. Moreover, a $\pi/2$ magnetic flux threads through each triangular plaquette in  clockwise and counter-clockwise directions for $\phi=0$ and $\phi=\pi$  due to the complex next-nearest-neighbor hopping in Eq.~(\ref{HamilefffinalSM21}). Such a nonzero flux leads to opposite particle transports along the odd and even lattice sites \cite{PhysRevLett.128.223903,PhysRevB.106.035425, PhysRevLett.122.023601}. Due to the larger loss for the even sites,   the particle transport along the odd sites are favored, and the backflow on even sites is suppressed. This leads to skin modes at right and left boundaries for $\phi=0$ and $\phi=\pi$, respectively. 

\begin{figure}[!tb]
	\centering
	\includegraphics[width=8.4cm]{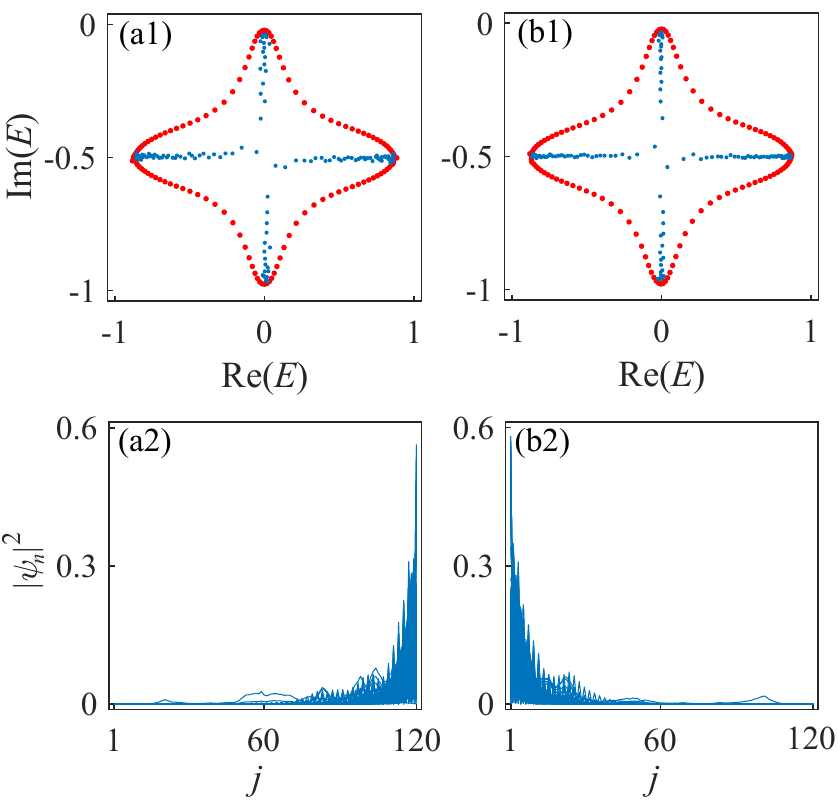}
	\caption{Complex quasienergy spectrum $E_n$ and corresponding density distributions $\abs{\psi_n}^2$ of the  Floquet Hamiltonian $\mathcal{H}_\textrm{F}$ in the presence of   onsite disordered potential   $\mathcal{H}_{\text{dis}}$ for (a1, a2) $\phi=0$, and (b1, b2) $\phi=\pi$. The red and blue dots indicate the quasienergies under PBC and OBC, respectively. The other parameters used are $\alpha=1.5$, $\omega=3\pi$, $\lambda/J=1$, $W/J=5$, and $L=120$.}\label{DisorderEffect} 
\end{figure}

\begin{figure}[!tb]
	\centering
	\includegraphics[width=9cm]{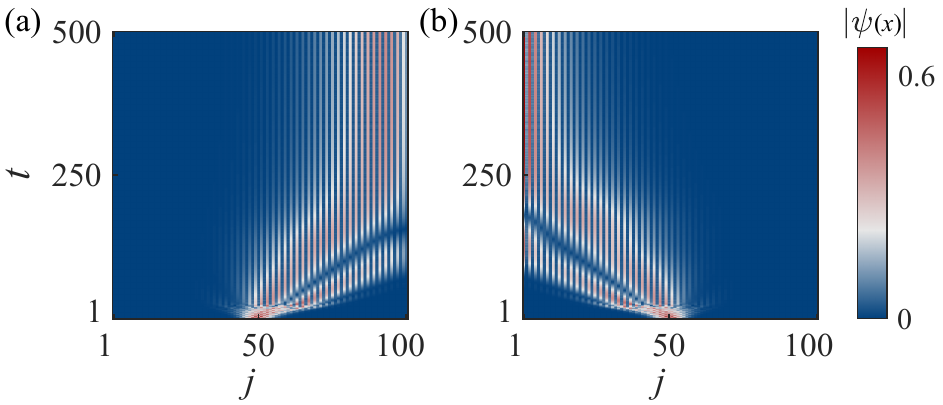}
	\caption{Dynamical localization of the periodically-driven dissipative system with (a) $\phi=0$, and (b) $\phi=\pi$, respectively. The initial states are chosen as  the Gaussian wavepacket $|\psi_0\rangle = \left[\psi_0(1),\psi_0(2),\dots,\psi_0(L)\right]^T$ with $\psi_0(j) =  \exp[-(j-j_0)^2/2\sigma^2]/\mathcal{N}$ centered at the site $j_0 = L/2$. Here we choose the width of the wavepacket $\sigma=5$. The wavepackets at every time have been renormalized.  The other parameters used are $\alpha=1.5$, $\omega=3\pi$, $\lambda/J=1$, and $L=100$. }\label{TimeEvolution} 
\end{figure}

\begin{figure}[!tb]
	\centering
	\includegraphics[width=8.6cm]{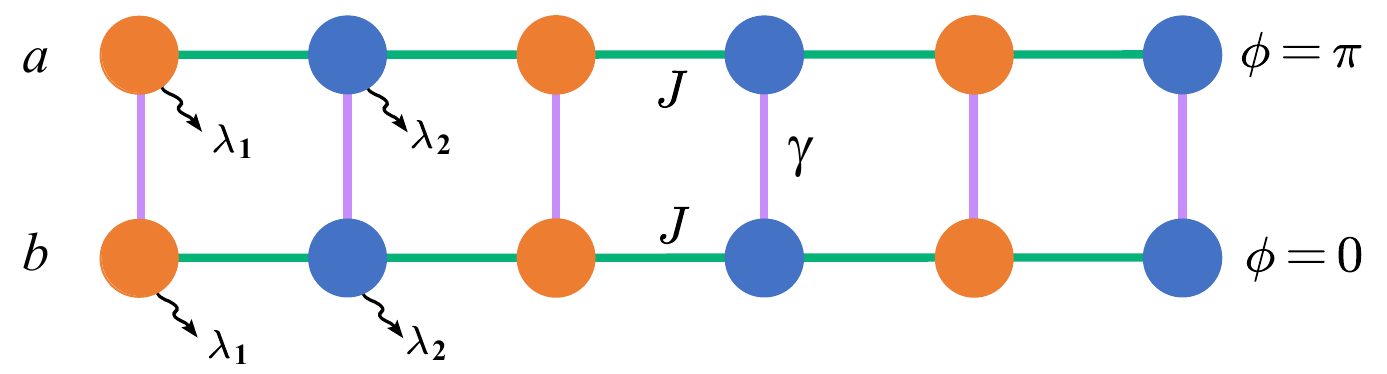}
	\caption{Schematic showing the lattice ladder of ultracold atoms for realizing critical NHSE. For each leg $a$ or $b$ of the ladder, the optical lattice has the same structure as the one in Fig.~\ref{FigScheme}, which is periodically modulated and staggered with alternating loss $\lambda_1$ and $\lambda_2$ with $\lambda = \lambda_2 - \lambda_1$. Along the rung,   driving phases are   set as $\phi=0$ and $\phi = \pi$, respectively. $\gamma$ is the hopping strength along the rung.}\label{Ladder} 
\end{figure}

\begin{figure*}[!tb]
	\centering
	\includegraphics[width=15cm]{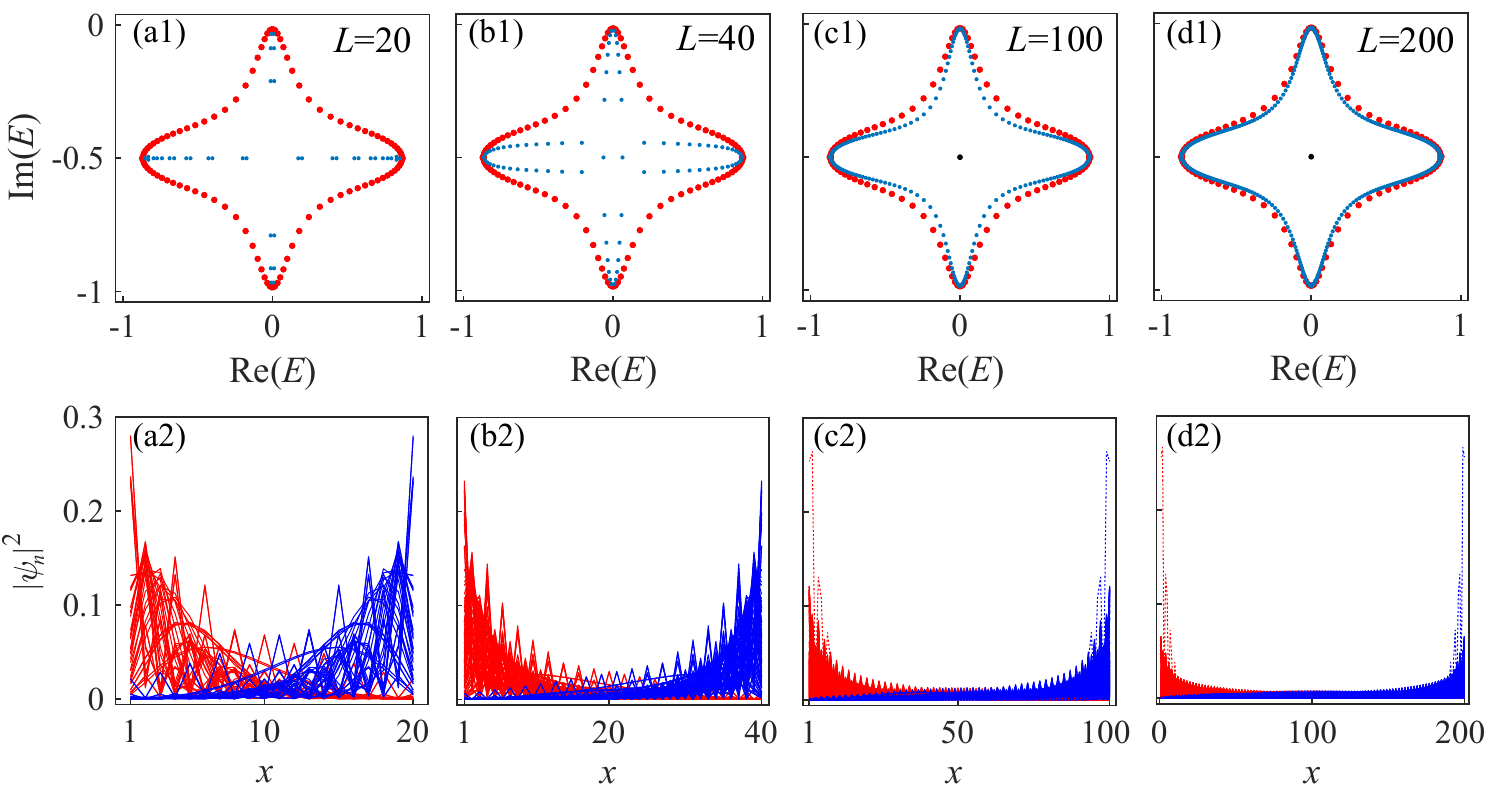}
	\caption{ Complex quasienergy spectrum $E_n$ and corresponding density distributions $\abs{\psi_n}^2$ of the  Floquet Hamiltonian $\mathcal{H}^{\textrm{F}}_\textrm{L}$ of the driving ladder for different size from  $L=20$ to $200$ . The red and blue dots indicate the quasienergies under PBC and OBC, respectively. The red and blue lines represent the density distributions $\abs{\psi_n}^2$ of legs $a$ and $b$, respectively. As the size $L$ increases, the topological in-gap   degenerate states (black dots) with zero real parts of complex eigenenergies appear  in (c1,d1), and the corresponding state distributions are shown in (c2,d2) with dashed curves.	The  parameters used are $\alpha = 1.5$, $\omega = 3\pi$, $\lambda/J = 1.0$, and $\gamma/J=0.005$. }\label{CriticalSkinEffect}
\end{figure*}

\subsection{Robustness Against Disorder}

The intrinsic topological nature of the NHSE in periodically-driven dissipative ultracold-atom system indicates its robust against local disorders. Here, we consider to introduce the onsite disordered potential into the periodically-driven dissipative system with  the system Hamiltonian reading 
\begin{align}\label{HamilDisorder}
\mathcal{H}_\textrm{tot}(t) = \mathcal{H}(t)  + 	\mathcal{H}_\textrm{dis} = \mathcal{H}(t) + \sum_{j=1}^{L} U_j n_{j}.
\end{align}
where $U_j$ denotes the onsite random potential, uniformly sampled in  $\in [-W/2,~W/2]$ with $W$ denoting the disorder strength. 

Figure \ref{DisorderEffect} shows the complex quasienergy spectra under OBC and PBC, and the corresponding density distributions  of the  Floquet Hamiltonian $\mathcal{H}_\textrm{F}$ in the presence of the  onsite random potential with the disorder strength $W=5$ for $\phi=0$  and  $\phi=\pi$. The PBC eigenenergies still form a point gap, encircling all the OBC eigenvalues. All the eigenstates are localized at the right (left) boundary for for $\phi=0$ and  $\phi=\pi$, respectively. Therefore, the NHSEs remain in spite of strong disorder due to the   intrinsic non-Hermitian topology of the periodically-driven dissipative ultracold-atom system considered here.

\subsection{Dynamical Localization}

The NHSE can be manifested by studying the dynamical evolution.  We consider the initial states as the Gaussian wavepacket $|\psi_0\rangle = \left[\psi_0(1),\psi_0(2),\dots,\psi_0(L)\right]^T$ with $\psi_0(j) =  \exp[-(j-j_0)^2/2\sigma^2]/\mathcal{N}$ centered at the site $j_0 = L/2$, where $\mathcal{N}$ is the normalization constant, and $\sigma$   denotes the width of the wavepacket. The wavefunction at time $t$ can be obtained by numerically calculating $|\psi(t)\rangle = \mathcal{T} \exp(-i \int_{0}^{t} \mathcal{H}(\tau) d\tau) |\psi_0\rangle$.

We calculate the time-dependent density distributions of the periodically-driven dissipative ultracold-atom system    for   $\phi=0$  and  $\phi=\pi$, as shown in Fig.~\ref{TimeEvolution}. Due to the nonreciprocal nature of the considered system, as analyzed above, the wavepacket is finally localized towards the right and left boundary for $\phi=0$  and  $\phi=\pi$, respectively.

\section{Floquet-Induced Critical Non-Hermitian Skin Effect}

The NHSE  can induce a novel critical behavior without its Hermitian counterpart, dubbed the critical NHSE \cite{Li2020}. That is, when two  non-Hermitian systems with different skin lengths are coupled together,  the energy spectrum  discontinuously jump across a critical point  in the thermodynamic limit. The critical skin effects have been studied in the static systems \cite{Li2020,PhysRevB.104.165117,PhysRevResearch.4.013243,PhysRevB.107.155430}. Here, we will show such a critical behavior can be also observed in the driving system.

To study Floquet-induced critical NHSE, we consider the quasi-one-dimensional ladder of ultracold atoms, as shown in Fig.~\ref{Ladder}. For each leg $a$ or $b$ of the ladder, the optical lattice has the same structure as the one in Fig.~\ref{FigScheme}, which is periodically modulated and staggered with alternating loss $\lambda_1$ and $\lambda_2$. Along the rung, driving phases are   set as $\phi=0$ and $\phi = \pi$, respectively. The Hamiltonian of the driving ladder is written in the rotating frame of reference (see Appendix \ref{Appendix_A})  as
\begin{align}\label{ladder}
	\mathcal{H}_\textrm{L}(t) =  	& -\sum_{j=1}^{L/2}  \left(J^\ast_- a^\dagger_{2j} a_{2j-1} + J^\ast_+ b^\dagger_{2j} b_{2j-1} + \text{H.c.}\right) \notag \\	
	& -\sum_{j=1}^{L/2-1}  \left(J_+ a^\dagger_{2j+1} a_{2j}  + J_- b^\dagger_{2j+1} b_{2j} + \text{H.c.}\right) \notag \\	
	& - \sum_{j=1}^{L/2} i \lambda \left(a_{2j}^\dagger a_{2j} + b_{2j}^\dagger b_2j \right) \notag \\	
	&-\sum_{j=1}^{L} \gamma \left(a_j^\dagger b_j + \text{H.c.} \right) ,
\end{align}
where $J_{\pm} = J e^{i\alpha\cos(\omega t \pm \frac{\pi}{4})}$ represents the hopping strength for the leg $a$ and $b$, $\lambda$ indicates onsite dissipation with $\lambda = \lambda_2 - \lambda_1$, and $\gamma$ is the hopping strength along the rung. The Floquet Hamiltonian $\mathcal{H}^{\textrm{F}}_\textrm{L}$ of the ladder is defined as $U(T)   =     \mathcal{T} e^{-i\int_{0}^{T} \mathcal{H}_\textrm{L}(t') dt'} = e^{-i \mathcal{H}^{\textrm{F}}_\textrm{L} T }$.

Figure \ref{CriticalSkinEffect} shows the complex quasienergy spectrum  and corresponding density distributions  of  $\mathcal{H}^{\textrm{F}}_\textrm{L}$ for different sizes $L$. The PBC spectrum (red dots) is not sensitive to the system size, while the OBC spectrum (blue dots) changes remarkably as the size $L$ of the  coupled chain increases from $L=20$ to $200$ in spite of the weak coupling strength between two chains with $\gamma/J = 0.005$. For the small size $L=20$, the OBC spectrum of the coupled chains mostly resembles the one of respective single chain in Fig.~\ref{energydensity}(d1,e1). As the size $L$ increases,  the OBC spectrum  changes, and the bulk OBC spectrum  approaches the PBC spectrum of the coupled chain at the large size $L=200$ [see Fig.~\ref{CriticalSkinEffect}(d1)]. Moreover, the energy  spectrum in the large-size  coupled chains under OBC discontinuously jumps across the critical point $\gamma/J = 0$. Therefore, the critical skin effect occurs in the periodically-driven ladder. 

Mostly remarkably, the critical skin effect can be manifested by the size-dependent topological phase crossovers. For $\gamma/J = 0$, the system is decoupled into two topologically trivial chains. While, for the nonzero coupling strength $\gamma/J \neq 0$, the momentum-space Hamiltonian $ H_\textrm{L}(k,t)$ in Eq.~(\ref{ladder}) respects chiral symmetry, which supports topological boundary modes (see details in Appendix \ref{Appendix_C}). As shown in Fig.~\ref{CriticalSkinEffect}(a1-d1), there exist no in-gap states for the small size of the ladder [see Fig.~\ref{CriticalSkinEffect}(a1,b1)]. However, as the size increases, topological degenerate in-gap states appear [see black dots in Fig.~\ref{CriticalSkinEffect}(c1,d1)]. These in-gap states are localized at the system's boundaries [see Fig.~\ref{CriticalSkinEffect}(c2, d2)]. In experiments,   we can observe the critical non-Hermitian skin effects by  directly  detecting the topological phase crossovers of the size-dependent in-gap boundary state. In ultracold atoms \cite{PhysRevLett.110.200406,Goldman2013, Chomaz2015,PhysRevLett.120.060402,arXiv:2304.01980}, the interface boundary between topologically trivial and nontrivial chains can be created by introducing a much large potential step (see details in Appendix \ref{Appendix_C}).

\section{Conclusion}

We have shown theoretically that the NHSE can occur in a periodically-driven  ultracold atoms in the presence of the staggered atomic loss. The NHSE is characterized by the winding number, which is quite robust against the disorder. The underlying mechanism governing  periodic-drive-induced NHSE is provided by considering Floquet-Magnus expansion in the high-frequency approximation. Moreover, we propose to observe the critical skin effect with the appearance of size-dependent topological in-gap modes by coupling two periodically-driven chains with different driving phases.  Our approach can be easily implemented in ultracold atoms based on Floquet engineering. Floquet engineering  has emerged as a powerful experimental method for the realization of novel quantum systems in  ultracold-atom systems. Our work paves the way for further studies of the NHSE  and its interplay with many-body statistics and interactions in ultracold-atom systems.

\begin{acknowledgments}
	T.L. acknowledges the support from Introduced Innovative Team Project of Guangdong Pearl River Talents Program (Grant
	No. 2021ZT09Z109), the Fundamental Research Funds for the Central Universities (Grant No.~2023ZYGXZR020),  National Natural
	Science Foundation of China (Grant No. 12274142), and the Startup Grant of South China University of Technology (Grant No.~20210012).
\end{acknowledgments}

\appendix
\section{Realization of Floquet Hamiltonian in Ultracold atoms}
\label{Appendix_AA}

\begin{figure*}[!tb]
	\centering
	\includegraphics[width=18cm]{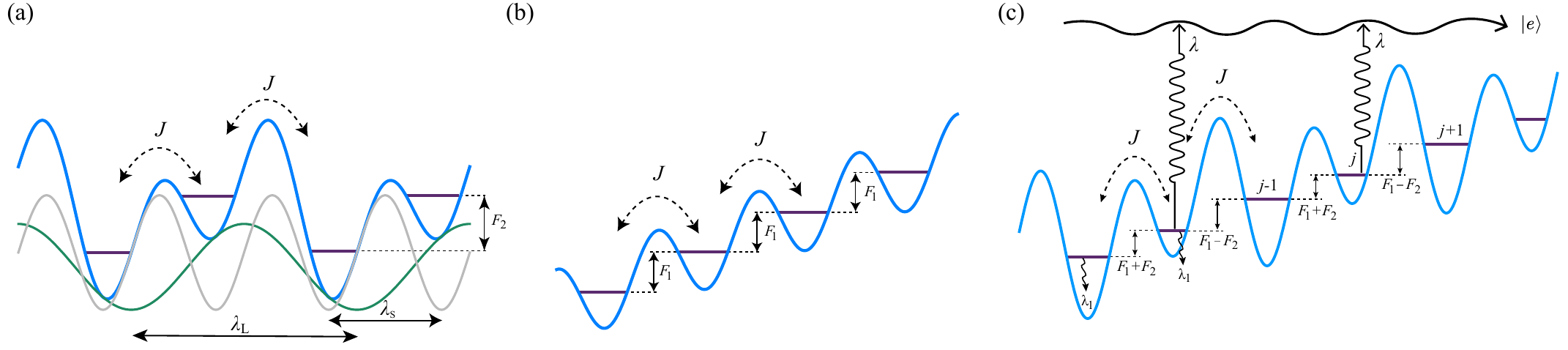}
	\caption{Schematic showing the dynamical superlattice potential to  realize the periodically-driven Hamiltonian $\mathcal{H}_\textrm{p}(t) = \mathcal{H}_0 + \mathcal{H}_\textrm{d}(t)$. The periodically-modulated potential $\mathcal{H}_\textrm{d}(t)$ consists of the time-dependent staggered onsite potential in (a), and time-dependent tiled onsite potential in (b). (a) The  one-dimensional dynamical optical superlattice with the time-dependent staggered onsite potential is created by superimposing a stationary lattice and a dynamical one with different periodicities. (b) An additional time-periodic modulation of the lattice position is superimposed to create the  time-dependent tilted potential. (c) When the atoms are prepared in the first-band state, the staggered  dissipation rates $\lambda_1$ and $\lambda_2 = \lambda_1 + \lambda$ at odd and even sites in $\mathcal{H}_0$ are realized via exciting  the  state   in the first band at even sites to the higher-level state $|e\rangle$, where the additional dissipation $\lambda$ is controlled by the intensity of the external optical beam.}\label{FigSchemeExp}
\end{figure*}

In this Appendix, we provide some details to realize the dissipative  Floquet Hamiltonian $\mathcal{H}_\textrm{p}(t) = \mathcal{H}_0 + \mathcal{H}_\textrm{d}(t)$ in Eqs.~(\ref{single1}) and (\ref{periodically}), which consist of the periodically-modulated staggered onsite potential and tilted potential and the staggered  dissipation rates $\lambda_1$ and $\lambda_2 = \lambda_1 + \lambda$ at odd and even sites.

As shown in Fig.~\ref{FigSchemeExp}(a),  in the experimental setting, the  one-dimensional dynamical optical superlattice (along the $x$ direction) in the presence of the periodically-modulated staggered onsite potential can be created by superimposing a stationary lattice (i.e., the short lattice) with a period of $\lambda_S$ and a dynamical interferometric lattice (i.e., the long lattice) with a period of $\lambda_L = 2 \lambda_S$ \cite{Lohse2015,Nakajima2016,Nakajima2021}. The phase of the dynamical interferometric lattice is periodically driven in time and controlled by a Michelson interferometer. The strong confinement along the $y$ and $z$ directions provided by other optical lattices.  The dynamical superlattice potential is written as
\begin{align}\label{Vxt}
	V_1(x,y,z, t) = & -V_S(t) \cos^2\left(k_S x\right) \nonumber \\ & -V_L(t) \cos^2\left[k_S x/2 - \varphi(t)\right],
\end{align}
where $V_S$ and $V_L$ is the depth of the short and long lattice controlled by the respective laser powers, $k_S = 2\pi/\lambda_S$ is the wave vector and $\varphi(t)$ is the phase difference between the two lattice controlled by changing the optical path difference between the two interfering beams \cite{Lohse2015,Nakajima2016,Nakajima2021}. The ultracold atomic gases are loaded into  this dynamically controlled optical superlattice, and prepared in the first band.  The band gap between the first and the higher-level bands are fairly large at each well, we can focus on only the first   band \cite{Lohse2015,Nakajima2016,Nakajima2021}. Then, in the tight-binding limit, we can achieve periodically-modulated staggered onsite potential in the second term of the right side of the Hamiltonian $\mathcal{H}_\textrm{d}(t)$ in Eq.~(\ref{periodically}).

Secondly, in order to construct the periodically-modulated tilted potential, we superimpose an additional drive consisting of a time-periodic modulation of the lattice position $x$ with the frequency $\omega$, as shown in Fig.~\ref{FigSchemeExp}(b). The time-periodic modulation  of the lattice site has the form 
\begin{align}\label{Vxt2}
	V_2(x,y,z, t) = -V(x-A\cos(\omega t + \phi_c)),
\end{align}
where $A$ is the modulation amplitude, and $\phi_c$ is phase. Such a  tilted potential under periodical-modulation (i.e., the time-dependent force) has   been experimentally realized in ultracold-atom systems  \cite{PhysRevA.96.053602, Grg2019}. Then, in the tight-binding limit,  we can achieve periodically-modulated tilted potential in the first term of the right side of in the Hamiltonian $\mathcal{H}_\textrm{d}(t)$ in Eq.~(\ref{periodically}).

Finally, when the atoms are prepared in the first-band state, the staggered  dissipation rates $\lambda_1$ and $\lambda_2 = \lambda_1 + \lambda$ at odd and even sites in $\mathcal{H}_0$ are realized via exciting  the  state   in the first band at even sites to the higher-level state $|e\rangle$, where the additional dissipation $\lambda$ is controlled by the intensity of the external optical beam [see Fig.~\ref{FigSchemeExp}(c)].  In order to selectively excite the atoms at even sites, we can utilize the ground-state hyperfine states in the combination of the Raman-assisted hopping between adjacent sites.  To be specific, we consider  ultracold atoms   with two internal states, e.g., spin up and spin down. The spin-up states at the odd sites are hopped to spin-down states at  the even sites using Raman-assisted hopping. Then, the staggered dissipation can be realized by exciting all the spin-down states at the even sites to higher-level states.  Note that the loss control using different internal states in optical lattice of ultracold atoms have been experimentally reported \cite{PhysRevLett.129.070401,Li2019,Ren2022,arXiv:2311.07931}.

\section{Unitary transformation of Floquet Hamiltonian}
\label{Appendix_A}

As shown in the main text, we construct the periodically-driven Hamiltonian $\mathcal{H}_\textrm{p}(t) = \mathcal{H}_0 + \mathcal{H}_\textrm{d}(t)$, with
\begin{align}\label{single1SM}
	\mathcal{H}_0 = -J \sum_{j=1}^{L-1}   \left(c^\dagger_{j+1} c_{j} + \text{H.c.}\right)   - i \lambda \sum_{j=1}^{L/2} n_{2j},
\end{align}
\begin{align}\label{periodicallySM}
	\mathcal{H}_\textrm{d}(t) = \sum_{j=1}^{L} F\left[ \cos(\omega t+\phi) j   +  \sin(\omega t)   \frac{3+(-1)^{j}}{2} \right]  n_{j},
\end{align}
where, without loss of generality, we set $\lambda_1 = 0$.

After performing a unitary transformation $\mathcal{H}(t) = \mathcal{U}\mathcal{H}_\textrm{p}(t)\mathcal{U}^\dagger - i \, \mathcal{U} \partial_t \mathcal{U}^\dagger $,  with $\mathcal{U}(t)$ being written as
\begin{align}\label{utransofromation}
	\mathcal{U}(t) = e^{i\left(\frac{F}{\omega} \sin(\omega t+\phi) \sum_{j}^{L}j n_{j} - \frac{F}{\omega} \cos(\omega t) \sum_{j}^{L} \frac{3+(-1)^{j}}{2} n_{j}\right)},
\end{align}
we achieve the periodically-driven Hamiltonian  as
\begin{align}\label{1DFmodel}
	\mathcal{H}(t) =  &  \sum_{j=1}^{L/2} \left[\left(u c^\dagger_{2j} c_{2j-1}  + \text{H.c.}\right) -i \lambda  n_{2j} \right] \nonumber \\
	&  - \sum_{j=1}^{L/2-1}  \left(  v c^\dagger_{2j+1} c_{2j} + \text{H.c.}\right) ,
\end{align}
where 
\begin{align}\label{1DFmode21}
	u = -J e^{i \sqrt{2}\alpha \cos(\frac{\phi}{2} + \frac{\pi}{4}) \sin(\omega t + \frac{\phi}{2} -\frac{\pi}{4})},
\end{align}
\begin{align}\label{1DFmode22}
	v =- J e^{i \sqrt{2}\alpha \sin(\frac{\phi}{2} + \frac{\pi}{4}) \cos(\omega t +\frac{\phi}{2}-\frac{\pi}{4})},
\end{align}
with $\alpha = \sqrt{2} F /\omega$.


\section{ Effective Hamiltonian via Floquet-Magnus Expansion}
\label{Appendix_B}

In this appendix, we provide some details to obtain the time-independent effective Hamiltonian $\mathcal{H}_\textrm{eff}$ by
applying the Floquet-Magnus expansion \cite{Bukov2015, Eckardt2015, PhysRevB.93.144307} in the high-frequency approximation ($\omega \gg J, F$). Up to the first-order approximation, the effective Hamiltonian is written as 
\begin{align}\label{Magnus}
	\mathcal{H}_{\text{eff}} = \sum_{\mu = 0,1}  \mathcal{H}_\textrm{eff}^{(\mu)},
\end{align}
where
\begin{align}\label{expansion12}
	\mathcal{H}_\text{eff}^{(0)} = \mathcal{H}_0, ~~~ \mathcal{H}_\text{eff}^{(1)} = \sum_{m\neq0} \frac{[\mathcal{H}_{-m},~\mathcal{H}_{m}]}{2 m \omega},
\end{align}
with $\mathcal{H}_m = T^{-1} \int_{0}^{T} \mathcal{H}(t)e^{im\omega t} dt $. 

We first derive the effective high-frequency Hamiltonian for the real-space model $\mathcal{H}(t)$ in Eq.~(\ref{1Dmodel}) of the main text. For $m \neq 0 $ and $\phi = 0 , \pi$, $\mathcal{H}_{m}$ reads
\begin{align}\label{HamilmSM2}
	\mathcal{H}_{m} = &  \sum_{j} \left[P_m(\frac{3\pi}{4}) c^\dagger_{2j} c_{2j-1} +  P_m(-\frac{\pi}{4}) c^\dagger_{2j-1} c_{2j}\right] \notag \\
	& + \sum_{j} \left[P_m(\frac{\pi}{4}) c^\dagger_{2j+1} c_{2j} + P_m(-\frac{3\pi}{4}) c^\dagger_{2j} c_{2j+1}\right],
\end{align}
where
\begin{align}\label{PmPsi}
	P_{m}(\Phi) = -J i^{-m} J_{-m}(\alpha)e^{ims\Phi},
\end{align}
with $J_m(\alpha)$ being the Bessel function of the first kind, and $s = 1 ~(-1)$ corresponding to $\phi = 0 ~(\pi)$.

\begin{figure}[!tb]
	\centering
	\includegraphics[width=8.4cm]{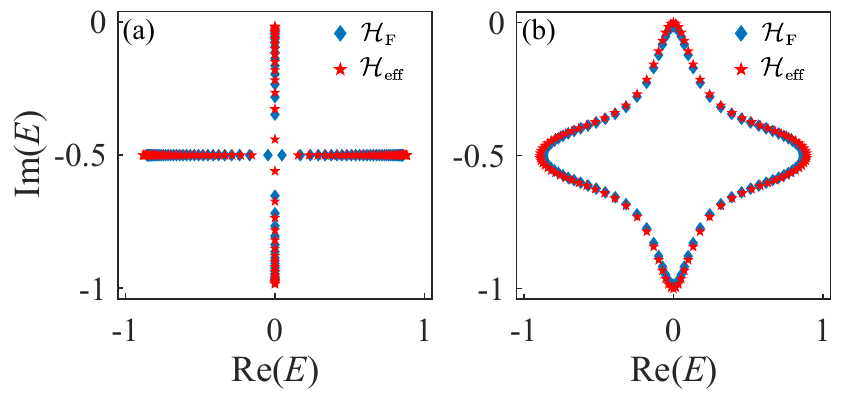}
	\caption{Complex quasienergy spectrum $E_n$ calculated using the Floquet Hamiltonian $\mathcal{H}_\textrm{F}$ (blue diamond dots) and effective Hamiltonian $\mathcal{H}_\textrm{eff}$ (red star dots) in the high-frequency limit under OBC (a) and PBC (b). The parameters used are $\alpha = 1.5$, $\omega=3\pi$, $\lambda/J=1$, and $L=100$.}\label{ExpansionFig1} 
\end{figure}

According to Eqs.~(\ref{Magnus}-\ref{HamilmSM2}), we  obtain the effective time-independent Hamiltonian $\mathcal{H}_\textrm{eff}$  of $\mathcal{H}(t)$ as
\begin{align}\label{HamilefffinalSM2}
	\mathcal{H}_\textrm{eff} = &- \sum_{j} \left[JJ_{0}(\alpha)c^\dagger_{2j} c_{2j-1} + JJ_{0}(\alpha) c^\dagger_{2j+1} c_{2j} + \textrm{H.c.} \right]   \notag \\
	& - \sum_{j=} \left[i s\frac{2J^2 J_{-1}^2(\alpha)}{\omega} c^\dagger_{2j+1} c_{2j-1} + \textrm{H.c.}\right]    \notag \\
	& + \sum_{j} \left[is\frac{2J^2 J_{-1}^2(\alpha)}{\omega} c^\dagger_{2j+2} c_{2j} + \textrm{H.c.}\right]  \notag \\
	& -i\lambda \sum_{j} c^\dagger_{2j} c_{2j},
\end{align}
where only the term with $m=1$ is considered in the high-frequency limit in Eq.~(\ref{expansion12}).

We now derive the effective high-frequency Hamiltonian for the momentum-space Hamiltonian $\mathcal{H}(k,t)$. We rewrite $\mathcal{H}(t)$ in Eq.~(\ref{1Dmodel})   as 
\begin{align}\label{1Dmodelre}
	\mathcal{H}(t) =  &   \sum_{j} \left[\left(u c^\dagger_{j, B} c_{j,A} + v c^\dagger_{j+1, A} c_{j, B} + \text{H.c.}\right) -i \lambda  n_{j, B} \right].
\end{align}
Then, the momentum-space Hamiltonian $\mathcal{H}(k,t)$ is written as 
\begin{align}\label{Hamildkt}
	\mathcal{H}(k,t) = & \sum_{k} \left[\left(u^\ast+v e^{-ik} \right) c^\dagger_{k,A} c_{k,B} + \text{H.c.}\right] \nonumber \\ &
	-i \lambda \sum_{k} c^\dagger_{k,B} c_{k,B},
\end{align}
where $A$ and $B$ denote two sublattice sites in the unit cell due to the staggered loss.

For $m \neq 0 $ and $\phi = 0 , \pi$, $\mathcal{H}_{m}(k)$ reads  
\begin{align}\label{HamilmkSM3}
	\mathcal{H}_{m}(k) = \sum_{k} \left[Q_{m,1}(k) c^\dagger_{k,A} c_{k,B} + Q_{m,2}(k) c^\dagger_{k,B} c_{k,A} \right],
\end{align}
where
\begin{align}\label{Qm1}
	Q_{m,1}(k) = -J i^{-m} J_{-m}(\alpha) \left(e^{-ism\frac{\pi}{4}} + e^{ism\frac{\pi}{4}} e^{-ik} \right),
\end{align}
\begin{align}\label{Qm2}
	Q_{m,2}(k) = -J i^{-m} J_{-m}(\alpha) \left(e^{ism\frac{3\pi}{4}} +  e^{-ism\frac{3\pi}{4}} e^{ik}\right).
\end{align}

According to Eqs.~(\ref{Magnus}) and (\ref{HamilmkSM3}), we  obtain the effective time-independent Hamiltonian $\mathcal{H}_\textrm{eff}(k)$ of $\mathcal{H}(k,t)$ as
\begin{align}\label{HamileffkfinalSM}
	\mathcal{H}_\textrm{eff}(k) = & - \sum_{k} \left[JJ_{0}(\alpha) \left(1+e^{-i k}\right) c^\dagger_{k,A} c_{k,B} + \textrm{H.c.} \right] \notag \\
	& - s\frac{2J^2 J_{-1}^2(\alpha)}{\omega} \sum_{k} \sin(k) \left( c^\dagger_{k,A} c_{k,A} - c^\dagger_{k,B} c_{k,B} \right) \notag \\
	&   -i\lambda \sum_{k} c^\dagger_{k,B} c_{k,B},
\end{align}
with   only the term with $m=1$ is considered in the high-frequency limit in Eq.~(\ref{expansion12}).

In Fig.~\ref{ExpansionFig1}, we plot the OBC and PBC spectra, calculated using the Floquet Hamiltonian $\mathcal{H}_\textrm{F}$ (blue diamond dots) and effective Hamiltonian $\mathcal{H}_\textrm{eff}$ (red star dots). The effective Hamiltonian in the high-frequency limit shows a good approximation to the  Floquet Hamiltonian.

\begin{figure}[!b]
	\centering
	\includegraphics[width=8.6cm]{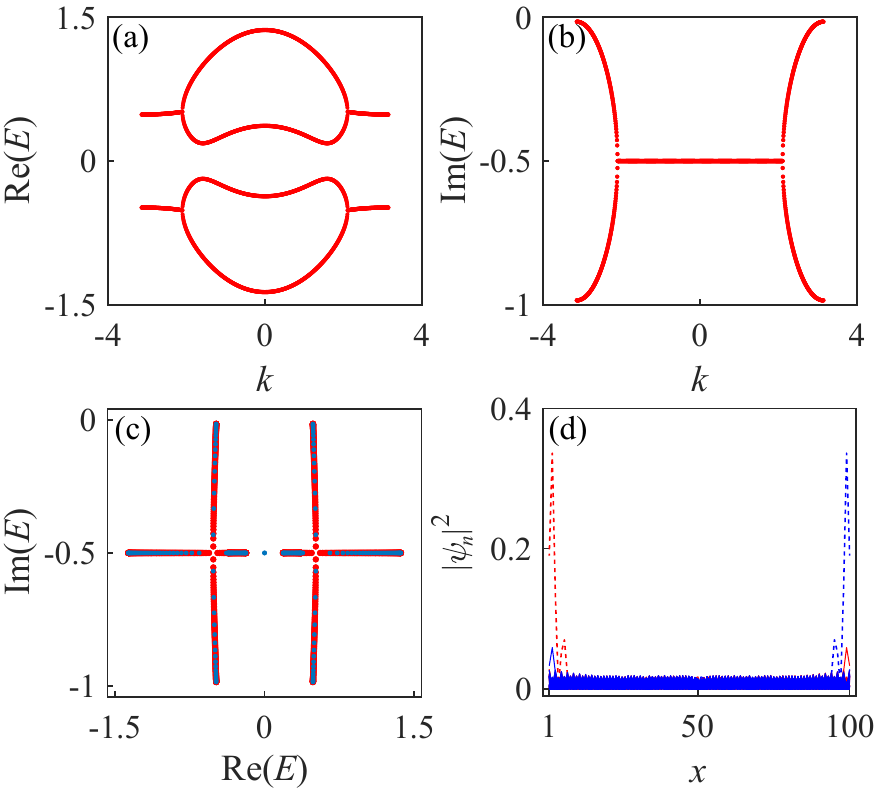}
	\caption{(a) Real part and (b) imaginary part of complex quasienergy spectrum of the Floquet Hamiltonian $\mathcal{H}_{\text{L}}^\textrm{F}(k)$ of $\mathcal{H}_{\text{L}}(k,t)$ in the momentum space. (c) Complex quasienergy spectrum $E_n$ of the ladder under PBC (red dots) and OBC (blue dots). (d) The corresponding density distribution of the ladder under OBC. The red and blue lines represent states in legs $a$ and $b$, respectively. The dashed lines  denote  the topological in-gap boundary states. The parameters are chosen as $\alpha=1.5$, $\omega=3\pi$, $\lambda/J=1$, $\gamma/J=0.5$, and $L=100$.}\label{TopologicalInsulators} 
\end{figure}

\section{Topological Phase of Periodically-Driven Ladder}
\label{Appendix_C}

In this appendix, we show more details of topological phases of the ladder system (see Fig.~\ref{Ladder}), which are protected by the energy gap and chiral symmetry. The momentum-space Hamiltonian of the ladder is written as $\mathcal{H}_\textrm{L}(k,t) = \sum_{k} \Psi_k^\dagger H_\textrm{L}(k,t) \Psi_k$, with $\Psi_k = (a_{k,A},  a_{k,B}, b_{k,A},  b_{k,B})^T$, where $A$ and $B$  denote    sublattices along each leg, as shown by yellow and blue filled circles in Fig.~\ref{Ladder}, and $	H_\textrm{L}(k,t)$ is
\begin{widetext}
	\begin{align}\label{Hamladderk}
		H_\textrm{L}(k,t)  = \left(\begin{matrix}
			0 & -J_- - J_+ e^{-ik} & -\gamma & 0 \\
			-J^\ast_- - J^\ast_+ e^{ik} & -i \lambda & 0 & -\gamma \\
			-\gamma & 0 & 0 & -J_+ - J_- e^{-ik} \\
			0 & -\gamma & -J^\ast_+ - J^\ast_- e^{ik} & -i\lambda
		\end{matrix}\right).
	\end{align}
\end{widetext}
The Floquet Hamiltonian $H_\textrm{L}^{\textrm{F}} (k)$ of the ladder in momentum space is defined as $U(T)   =     \mathcal{T} e^{-i\int_{0}^{T} H_\textrm{L}(k,t')dt'} = e^{-i H_\textrm{L}^{\textrm{F}} (k) T }$.

The Hamiltonian $H_\textrm{L}(k,t)$ respects chiral symmetry $S H_\textrm{L}(k,t) S^{-1} = - H^\dagger_\textrm{L}(k,-t) $, where $S = \sigma_y \tau_z $, with Pauli matrices $\sigma_{0,x,y,z}$ and $\tau_{0, x,y,z}$   acting on the sublattices $(A,B)$ in each leg  and legs $(a,b)$ degrees of freedom, receptively. As shown in Fig.~\ref{TopologicalInsulators}(a,b), the Floquet Hamiltonian $H_\textrm{L}^{\textrm{F}} (k)$ has a line gap. According to topological classification of Floquet  non-Hermitian system \cite{PhysRevB.105.214305}, it supports a topological nontrivial phase.   Figure   \ref{TopologicalInsulators}(c,d) clearly shows the degenerate in-gap boundary modes.

\begin{figure*}[!tb]
	\centering
	\includegraphics[width=17cm]{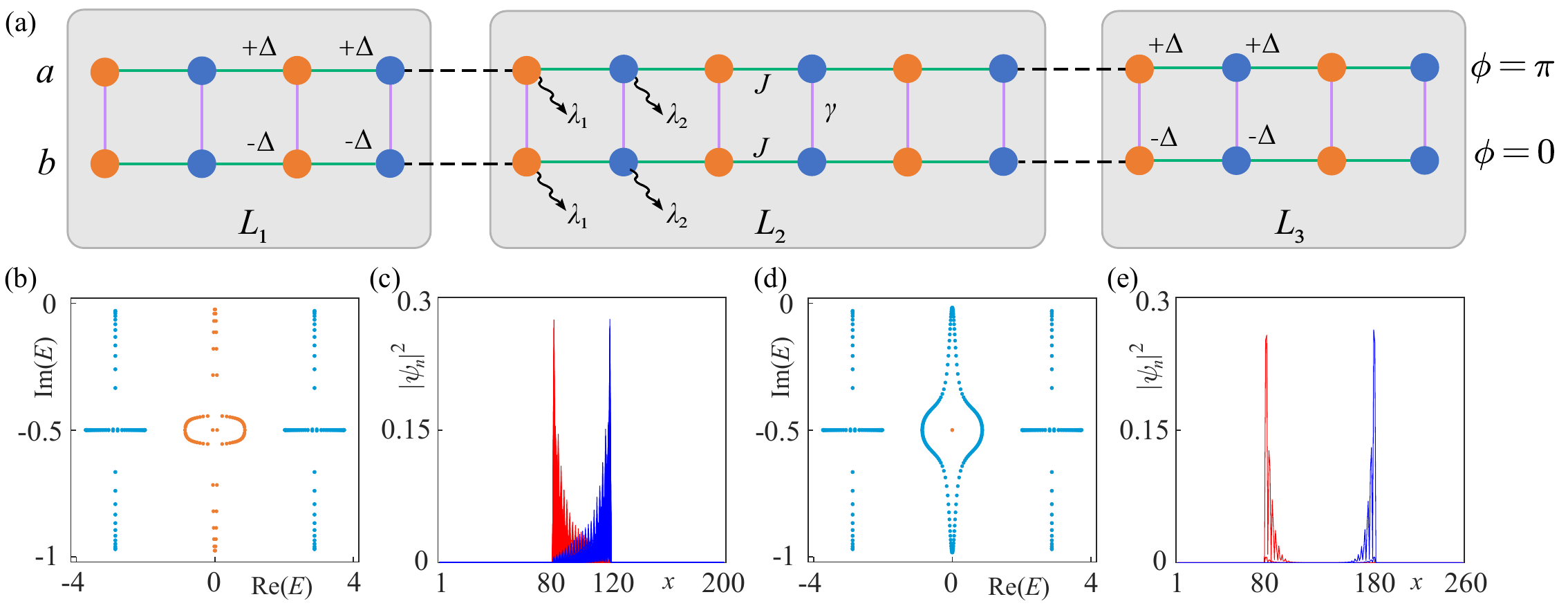}
	\caption{ (a) A middle ladder with the size $L_2$ is sandwiched by two ladders with lengths $L_1$ and $L_3$  in the presence of the much large onsite potential $ \Delta$ and $-\Delta$ in chains $a$ and $b$. (b) Complex-eigenenergy spectrum and (c) the corresponding    density distributions of states indicated by  red dots in (b) for $L_2 = 40$. (d) Complex-eigenenergy spectrum and (e) the density distributions of the topological in-gap states ($E=-0.5i$)  indicated by  red dots in (d) for  $L_2 = 100$. The  parameters used are $\alpha = 1.5$, $\omega = 3\pi$, $\lambda/J = 1.0$, and $\gamma/J=0.005$.}\label{FigCritialPotential}
\end{figure*}

In experiments,   we can observe the critical non-Hermitian skin effects by  directly  detecting the topological phase crossovers of the size-dependent in-gap boundary state. In ultracold atoms, the interface boundary between topologically trivial and nontrivial chains can be created by illuminating a selected area of	the lattice with a large optical potential \cite{PhysRevLett.110.200406,Goldman2013, Chomaz2015,PhysRevLett.120.060402,arXiv:2304.01980}. As shown in Fig.~\ref{FigCritialPotential}(a), the middle ladder with the size $L_2$ is sandwiched by two ladders with lengths $L_1$ and $L_3$  in the presence of the much large onsite potential $ \Delta$ and $-\Delta$ in chains $a$ and $b$.  The left and right ladders are topologically trivial for large $\Delta$ due to the broken chiral symmetry.  We calculate the complex-eigenenergy spectrum and   the corresponding    density distributions for $L_2 = 40$ [see Fig.~\ref{FigCritialPotential}(b,c)] and $L_2 = 40$ [see Fig.~\ref{FigCritialPotential}(d,e)]. The results indicate that the eigenenergy spectrum and state density distribution of the middle ladder are largely separated from those of the left and right ladders, and  have the same features as the single ladder under OBCs (see Fig.~\ref{CriticalSkinEffect})  for large $\Delta$.

\end{document}